\documentclass[review, 3p]{elsarticle}

\usepackage{lineno}
\modulolinenumbers[5]

\usepackage{float}
\usepackage[caption=false]{subfig}
\usepackage{amssymb}
\usepackage{array}
\usepackage{amsmath}
\usepackage[colorlinks=true, allcolors=black]{hyperref}
\usepackage{booktabs} 
\usepackage{arydshln} 
\usepackage{multirow}

\DeclareMathOperator{\softmax}{{\mathsf{softmax}}}

\journal{Journal of Computer Speech and Language}

\biboptions{authoryear}

\begin{document}

\begin{frontmatter}

\title{AV Taris: Online Audio-Visual Speech Recognition}

\author{George Sterpu and Naomi Harte}
\address{Sigmedia Lab, ADAPT Centre, Department of Electronic and Electrical Engineering,\\ School of Engineering, Trinity College Dublin, Ireland}

\begin{abstract}
In recent years, Automatic Speech Recognition (ASR) technology has approached human-level performance on conversational speech under relatively clean listening conditions. In more demanding situations involving distant microphones, overlapped speech, background noise, or natural dialogue structures, the ASR error rate is at least an order of magnitude higher. The visual modality of speech carries the potential to partially overcome these challenges and contribute to the sub-tasks of speaker diarisation, voice activity detection, and the recovery of the place of articulation, and can compensate for up to 15dB of noise on average. This article develops AV Taris, a fully differentiable neural network model capable of decoding audio-visual speech in real time. We achieve this by connecting two recently proposed models for audio-visual speech integration and online speech recognition, namely AV Align and Taris. We evaluate AV Taris under the same conditions as AV Align and Taris on one of the largest publicly available audio-visual speech datasets, LRS2. Our results show that AV Taris is superior to the audio-only variant of Taris, demonstrating the utility of the visual modality to speech recognition within the real time decoding framework defined by Taris. Compared to an equivalent Transformer-based AV Align model that takes advantage of full sentences without meeting the real-time requirement, we report an absolute degradation of approximately 3\% with AV Taris. As opposed to the more popular alternative for online speech recognition, namely the RNN Transducer, Taris offers a greatly simplified fully differentiable training pipeline. As a consequence, AV Taris has the potential to popularise the adoption of Audio-Visual Speech Recognition (AVSR) technology and overcome the inherent limitations of the audio modality in less optimal listening conditions. Our code is publicly available at https://github.com/georgesterpu/Taris.
\end{abstract}


\end{frontmatter}


\section{Introduction}

The next frontiers in computer speech technology include the capacity to enable natural conversations between humans and computers. An essential requirement from such technology is the automatic recognition of the spoken words with minimum latency. A fast and accurate decoding of speech by computers enables a more interactive communication with humans, creating the possibility for acknowledgements or interruptions to plead for clarifications. For example, we may be at a museum asking for directions, place an order at a restaurant, or check-in at a hotel lobby. In all these situations, it is important to guarantee a short response time to allow an interactive and natural conversation.

Real time, or online speech recognition is already possible with traditional speech models, or with neural network models based on the the RNN Transducer (RNN-T) architecture of \cite{Graves2012}. The former models, including the GMM-HMM, despite generalising well on relatively small datasets for contemporary standards, are known to display diminishing returns for an increasing amount of speech data compared to their neural counterparts.
As for the latter, \cite{Prabhavalkar2017} note in their comparative study that the inference in the RNN Transducer has the potential to be performed in a frame-synchronous, hence real time mode, if coupled with a unidirectional encoder. Their work only investigated bidirectional encoders to allow a more fair comparison to the attention model. Indeed, the RNN-T model has already been tested in a practical setting, \cite{Sainath2020} showing that the RNN-T is comparable in latency and accuracy with a conventional model for only a fraction of the size. However, a shortcoming of RNN-T is its inference complexity, where two separate modules, the \emph{Prediction} and \emph{Transcription} networks, dynamically alternate their turns depending on the current output label being either a blank or non-blank token. \cite{Wang2019} analyse the shortcomings of the RNN-T, finding that its dynamic programming training algorithm marginalises over a large number of alignment paths including many unreasonable ones, and report training difficulties. Furthermore, \cite{Battengerg2017} note that bridging the modelling assumptions between the RNN-T and attention models, particularly by equipping attention models with the monotonicity constraints of the RNN-T, is a promising avenue. These attention models, commonly known as sequence to sequence (seq2seq) or encoder-decoder architectures~\citep{Cho2014, Sutskever2014, Forcada1997, Kalchbrenner2013}, address the shortcomings of the RNN-T, but introduce a different computation paradigm that no longer enables online speech recognition. Such models were initially proposed in Machine Translation, and have not been fully optimised to the structure of the speech signal. Consequently, our aim in this work is to adapt sequence to sequence models for the online processing of speech utterances.

The remarkable progress in the age of deep learning has set the word error rates on a par with the human level performance on relatively clean, structured conversations~\citep{Saon2017}, and has enabled end-to-end automatic speech recognition, where the seq2seq neural network can exceed the performance of the traditional approaches~\citep{Chiu2018}. Despite this, an unresolved issue is reducing the latency from full utterances down to a few words. The seq2seq model conditions every target unit on the full unsegmented audio sentence, being predicated on the principle that a decoder drives the soft segmentation of the input during training. Because the first output token can only be emitted once the entire input sequence has been encoded, this sentence-level, or offline conditioning, is a fundamental barrier in decoding speech in real-time, or online, with a seq2seq network. It has been shown that, once convergence is reached, there are predominantly local relationships between the output tokens and the audio representations in speech~\citep{Chorowski2015, Chan2016}. Therefore, potentially incurring no loss in accuracy, an explicit local conditioning of the outputs on the inputs would break the offline limitation and reduce the algorithmic latency. The new challenge is to learn robust associations between input and output subsequences which stand for the same linguistic concepts. We argue this is a necessary inductive bias in speech recognition, as the task specification sets no limit on maximum sequence lengths, and the truncation of long sentences is already performed during the collection of speech datasets.

Humans learn to simultaneously segment and recognise fluent speech from the earliest stages of life~\citep{Jusczyk1995}. \cite{Cairns1994} describe the relationship between speech segmentation and recognition as a chicken-and-egg problem: segmenting units with meaning (e.g. words) from continuous speech posits the recognition of the unit, but the recognition of a unit presumes its \emph{a priori} segmentation. Unlike in text, there are no clear markers of the boundaries between the spoken words. Moreover, many words in a vocabulary represent the prefix of another longer word, such as \emph{car} in \emph{carbon}. \cite{Luce1986} estimated that, prior to listening to the last phone of a word from a lexicon of 20,000 words in the English language, less than 50\% of the most frequent words (up to 5 phonemes in length) are phonemically unique. Their study concluded that many words cannot be recognised in fluent speech until the initial segment of the following word is identified. Furthermore, \cite{Saffran1996} describe words as clusters of syllables characterised by a higher \emph{transitional probability} between the syllables within the word than of the syllable pairs occurring at boundary between two consecutive words. \cite{Johnson2001} explain that we integrate a set of acoustic, phonetic, prosodic, and statistical cues in order to segment words in fluent speech.
A natural pathway is to investigate whether explicitly introducing the ability to segment speech into \emph{word} units with a neural network represents a reasonable inductive bias to address the challenge decoding speech in real-time. This approach would take advantage of the monotonicity of speech, allows the network focus on local properties, and removes the offline conditioning.

In our prior work ~\cite{Sterpu2020c}, we investigated the research question of whether neural networks can learn to segment a spoken utterance by learning \emph{to count} the words therein. Different from the way humans acquire a language, we suspected that there is a strong relationship between learning the ability to count spoken words, and the ability to segment words in fluent speech. We used this conjecture to develop Taris, an online speech recognition system that estimates word boundaries, and subsequently eagerly decodes clusters of segments into words. Different from the related models, the network responsible for the analysis of the boundary cues in Taris is trained by minimising the difference between a word count estimated from the audio features and the ground truth value inferred from the text labels. For completeness reasons, we will review Taris in Section~\ref{sec:taris}.

In this article, our research question extends the analysis of the segmentation approach in Taris to multiple modalities. We want to examine if words can be counted more reliably from the audio-visual representations of speech, than from audio alone. \cite{Binnie1974} find that the visual modality of speech provides cues regarding the place of articulation. Furthermore, \cite{Summerfield1987} suggests that the visual modality may play a role in speech segmentation and voice activity detection. These properties are particularly important in noisy environments with overlapped speech, where we expect a superior speech recognition performance when integrating cues from multiple sources. \cite{Mitchel2010} found that the visual modality of speech does not improve the ability to segment an artificial language through statistical learning. They suspected that, under clean speech conditions, the boundary cues extracted from the audio modality were sufficiently robust and made the visual cues unnecessary. Later, \cite{Mitchel2014} investigated the role of visual cues for speech segmentation by creating an artificial language with limited statistical cues for the detection of word boundaries. They found the visual speech cues from talking faces to be beneficial to speech segmentation. They also ensured that the prosody cues inferred from vision did not contribute to segmentation by restricting the head nodding of the talking actors. The initial results of \cite{Tan2019} may be indicative of the fact that infants pay more attention to the mouth area when exposed to audio-visual speech stimuli. Taken together, this represents a strong motivation to investigate the potential of the visual modality to improve the robustness of Taris at decoding online.

To this end, we introduce \emph{AV Taris}, a multimodal Transformer-based system for online speech recognition that extends Taris and operates on audio-visual speech inputs. We make use of our previously proposed method AV Align~\citep{Sterpu2018b, Sterpu2020a} for fusing the two speech modalities. To enable the real time processing of the input sequences, we restrict the cross-modal alignment operation in AV Align to a window of fixed size. Subsequently, the network responsible with the prediction of the word count in Taris now has access to fused audio-visual speech cues instead of purely auditory ones. We find that AV Taris achieves a slightly higher error rate than the equivalent offline Audio-Visual Align Transformer model when trained and evaluated under the same conditions, but is considerably superior to an audio-only model. Our result supports the hypothesis that the visual modality plays a significant role in the segmentation of a spoken utterance. Furthermore, the design of AV Taris preserves the property of Taris to decode speech in real time, on audio-visual inputs. We describe AV Taris in Section~\ref{sec:avtaris}. The name \emph{Taris} echoes the misuse of the strong-weak syllable stress rule when learning to segment words by infants exposed to the phrase \mbox{\emph{the guitar is}}, discussed in \cite{Jusczyk1999}.

\section{Taris}
\label{sec:taris}

We begin this section by first reviewing the underlying approach Taris presented in \cite{Sterpu2020c}. Taris was originally designed to process the auditory modality of speech. In this work we will refer to it as \emph{Audio Taris}, and we will discuss it in Section~\ref{sec:audio_taris}. We will extend this model to additionally take advantage of the visual modality of speech, introducing \emph{AV Taris} in Section~\ref{sec:avtaris}. Since the multimodal extension is a windowed variant of our previously proposed method \emph{AV Align}~\cite{Sterpu2018b}, we will briefly summarise this method in Section~\ref{sec:avalign}. When referring to the overall approach based on word counting and segment-based real-time decoding, independent of the audio or audio-visual nature of the speech cues, we will only use the term \emph{Taris}.

\begin{figure}[t]
        \centering
        \subfloat[Full connectivity]{
        \includegraphics[width=0.25\linewidth]{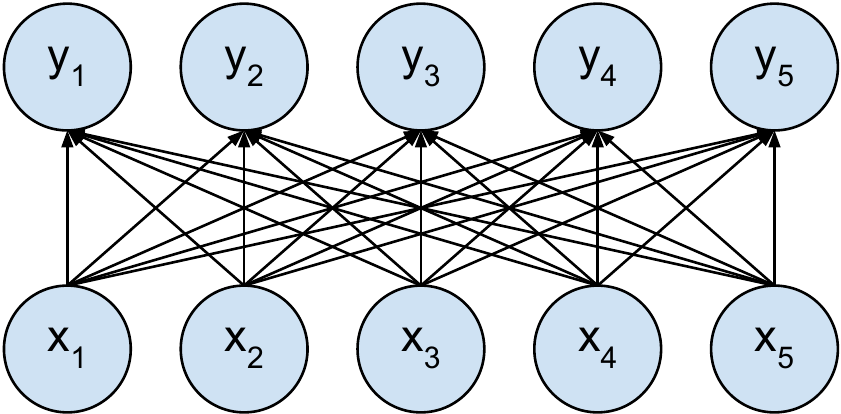}
           \label{fig:con:off}
        }
        \hspace{1cm}
        \subfloat[Causal connectivity]{
           \includegraphics[width=0.25\linewidth]{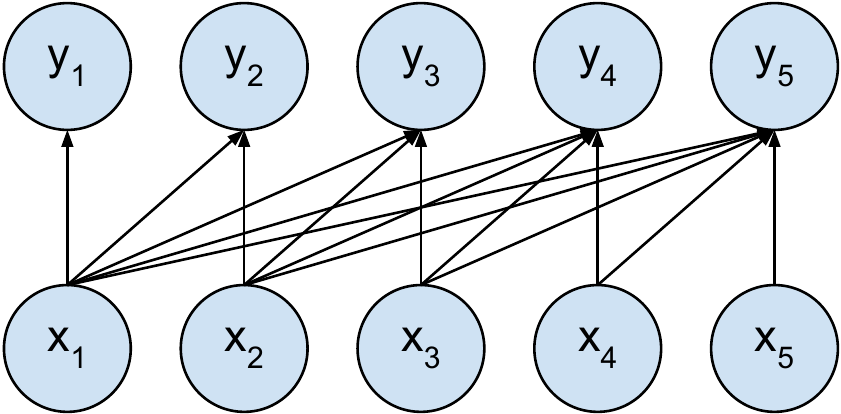}
           \label{fig:con:causal}
        }
        \hspace{1cm}
        \subfloat[Windowed connectivity with one context frame to left and right]{
        \includegraphics[width=0.25\linewidth]{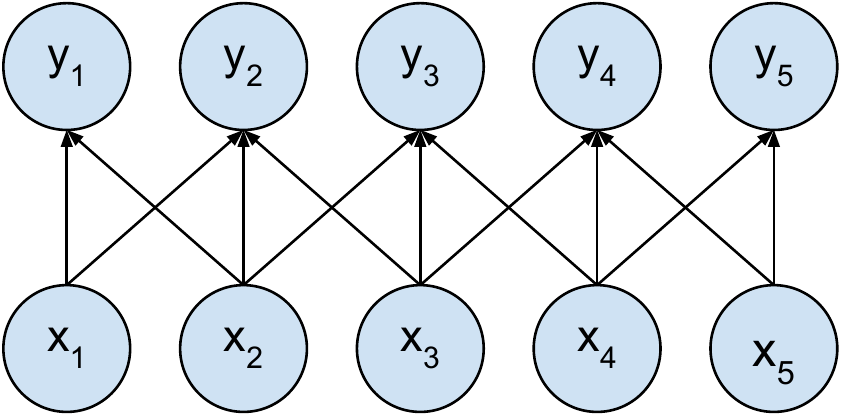}
           \label{fig:con:look}
        }
        \caption{Sequence-level conditioning strategies commonly used by neural network architectures in speech processing.}
        \label{fig:tmp5}
    \end{figure}

\subsection{Audio Taris}
\label{sec:audio_taris}

Taris takes as input a variable length sequence of audio vectors $\mathbf{a} = \{a_1, a_2, \ldots , a_{\mathbf{N}}\}$ and applies the Encoder stack of the Transformer model defined in~\cite{Vaswani2017}. A typical Transformer encoder has the sequence connectivity pattern illustrated in Figure~\ref{fig:con:off}, where each representation in a higher order layer is conditioned on the entire sequence of representations from the previous layer. Since we aim to control the latency of the encoding process for speech signals, Taris limits this conditioning to a fixed window centred on the sequence element with controlled look-back $e_{LB}$ and look-ahead $e_{LA}$ frames. Such an example for one look-ahead and look-back frames, respectively is displayed in Figure~\ref{fig:con:look}. Technically, we achieve the restraint of the range of the attention operation in the self-attention Encoder of the Transformer by masking those attention weights outside the allowed range with zeros. The Encode operation below denotes the masked variant of the original operation in the Transformer, and produces the audio representations $o_{A}$:
\begin{align}
    \mathbf{o_{A}}  = \mathsf{Encode}(\mathbf{a}, e_{LB}, e_{LA})
    \label{eq:a}
\end{align}

To obtain a soft, differentiable estimate of the word count from the encoder representations $o_{A}$, we start by applying a sigmoidal gating unit on each encoder output $o_{A_i}$ to obtain a scalar score for each frame:
\begin{align}
    \alpha_i & = \mathsf{sigmoid}(o_{A_i} W_G  + b_G) \label{eq:alphai} \\
    & \nonumber \mathsf{where}\ \mathsf{sigmoid}(x) = \frac{1}{1 + \exp(-x)}, W_{G} \in \mathbb{R}^{\mathrm{h\,x\,1}}, b_{G} \in \mathbb{R}^{\mathrm{1}}
\end{align}

Here, $h$ stands for the number of neurons defining the size of the hidden state of our model.

We assign to every single input frame $i$ a segment index $\hat{w_i}$ by taking the \emph{cumulative sum} of $\alpha$ and applying the \textit{floor} function on the output:
\begin{align}
    \hat{w_i} = \left\lfloor\sum_{j=1}^{i} \alpha_j  \right\rfloor
\end{align}

Namely, the first predicted segment is delimited by a cumulative sum of $\alpha$ between 0 and 1, the second segment by the same quantity between 1 and 2, and so on. 

During training, the Decoder stack receives the labelled grapheme sequence $\mathbf{y} = \{y_1, y_2, \ldots, y_{\mathbf{L}}\}$, made of English letters and the unique word delimiter SPACE. We assign to every grapheme $k$ a word index $w_k$ by leveraging the SPACE tokens in the labelled sequence:
\begin{align}
    w_k = \sum_{j=1}^{k} (y_j == \mathsf{SPACE}) \label{c4:eq:space}
\end{align}


Thus, whereas symbolic segmentation of speech uses a unique SPACE token to separate words, acoustic segmentation flags word boundaries by tracking the frame locations where the partial sum of the word counting signal $\alpha_i$ passes to the next integer value.

We modify the decoder-encoder connectivity of the Attention layer of~\cite{Vaswani2017} to allow our decoder to perform soft-alignment over a \emph{dynamic} window of segments estimated by the encoder. More precisely, we only allow those connections for which the following condition is met:
\begin{align}
    \mathbf{V} =  \widehat{W}_{ik} \leq (W_{ik} + d_{LA})\ \mathbf{and}\ \widehat{W}_{ik} \geq (W_{ik} - d_{LB})
    \label{eq:valid}
\end{align}

In equation~\eqref{eq:valid}, $d_{LA}$ and $d_{LB}$ denote the number of segments the decoder is allowed to look-ahead and look-back respectively. The $W$ and $\widehat{W}$ matrices are obtained from the $w$ and $\hat{w}$ arrays by applying the tile operation, which repeats one sequence for a number of times equal to the length of the other one.
\begin{align}
\widehat{W}_{ik} & = \hat{w_k}\\
W_{ik} & = w_k \\
\nonumber \forall i & \in [1, N] \\
\nonumber \forall k & \in [1, L]
\end{align}
For example, if we assume a 4-word sentence with $w = [000111222333]$ and $\hat{w} = [0123]$, tiling generates two matrices $W$ and $\widehat{W}$ of the same shape 12x4 by replicating the rows of w 4 times and the columns of $\hat{w}$ 12 times (and transposing). A commonly used implementation of this operation is \emph{numpy.tile}. We then extend the association between the indices of these matrices to support \emph{segment} look-back and look-ahead. More generally, $\mathbf{V}$ is a 2D matrix $\in \mathbb{R}^{\mathrm{N\,x\,L}}$ that defines the admissible connections between any decoder timestep and any encoder timestep, acting as a bias on the decoder-encoder attention. Setting $\mathbf{V}$ as a matrix of ones recovers the original Transformer model. The extension to 3D tensors that include the batch dimension is straightforward, offering Taris efficient minibatch training and inference. Note that the look-back and look-ahead parameters of the encoder and the decoder denote different units. Whereas $e_{LB}$ and $e_{LA}$ designate a number of input frames, $d_{LB}$ and $d_{LA}$ represent clusters of frames that we call \emph{segments}. The separation between \emph{segments} and \emph{words} is made on purpose to avoid making a strong claim regarding the nature of the speech segments at the end of system training.

The decoder implements a traditional character level auto-regressive language model that predicts the next grapheme in the sequence, conditioned on all the previous characters and the dynamic audio context vector $c_k$:
\begin{align}
    c_k & = \mathsf{Attention}(\mathtt{keys=}\mathsf{o_{A}},\mathtt{query=} o_{D_{k-1}}, \mathtt{mask=}V) \label{eq:tarisctx} \\
    o_{D_k} & = \mathsf{Decode}(y, c_k) \\
    p_k & \equiv P(y_{k} | c_k, y_{1 : k-1}) = \softmax(o_{D_k} W_{v} + b_v) \label{eq:vocab} \\
    \nonumber & \mathsf{where} \ W_{v} \in \mathbb{R}^{\mathrm{h\,x\,v}}, b_{v} \in \mathbb{R}^{\mathrm{v}} 
\end{align}

In equation~\eqref{eq:vocab}, $v$ is the alphabet size of 28 tokens representing the 26 English letters, space, and apostrophe.
We measure the difference between the estimated word sum $\Sigma \hat{w} = \sum_i \alpha_i$ and the true word count $|w| = \sum_k (y_k == \mathsf{SPACE})$ as:
\begin{align}
    Word\ Loss & = (|w| - \Sigma \hat{w})^2 \label{eq:wloss}
\end{align}

We define the training loss as:
\begin{align}
     CE\ Loss & = \frac{1}{L} \sum_k -y_k \log (p_k) \label{c4:eq:celoss} \\
    \mathsf{Loss} & = CE\ Loss + \lambda\ Word\ Loss
\end{align}

In all our experiments we used a scale factor $\lambda = 0.01$ found empirically. The self-attention connections of the Decoder are causal, describing an auto-regressive process.

    \subsection{AV Align}
    \label{sec:avalign}

    Since we are interested in a design that learns to count words from audio-visual speech cues as opposed to audio-only ones seen in the previous section, we need a multimodal fusion strategy that integrates the two streams at an early stage. There are two reasonable choices for the integration step. First, there is direct feature concatenation, which expects a similar sampling rate of the two speech modalities to enable the sequence-level concatenation operation. Second, there is our previously proposed method AV Align~\citep{Sterpu2018b, Sterpu2020a}, that applies an attention mechanism from the audio modality to the visual one, and subsequently obtains time-aligned fused audio-visual speech representations based on the dot product correlation score. To bypass the equal sampling rate limitation of direct feature concatenation, in this work we will focus on AV Align as our multimodal speech integration strategy.
    
    Given a variable length acoustic sentence $a = \{a_1, a_2, \ldots , a_N\}$ and its corresponding visual track $v = \{v_1, v_2, \ldots , v_M\}$ of length $M \neq N$, two separate stream encoders (e.g. Recurrent / Convolutional Neural Networks, Transformer) project the sequences onto higher order abstract representations ${o}_{A} = \{o_{A_1}, o_{A_2}, \ldots , o_{A_N}\}$ and ${o}_{V} = \{o_{V_1}, o_{V_2}, \ldots , o_{V_M}\}$):
\begin{align}
    o_{A} & = \mathsf{Encode} (a) \\
    o_{V} & = \mathsf{Encode} (v)
\end{align}

    AV Align obtains a fused sequence $o_{AV}$ where the audio representations $o_{A_i}$ are fused with their contextualised visual representations $c_{V_i}$ using an attention mechanism:
    \begin{align}
        \alpha_{ij} & =  \softmax_i(o_{A_i}^{T} \cdot o_{V_j}) \\
        & \mathsf{where} \ \softmax_i(\mathbf{x}) = \frac{\exp(x_i)}{\sum_j \exp(x_j)} \\
        c_{V_i}  & = \sum_{j=1}^{M} \alpha_{ij}\ \cdot o_{{V}_j}
    \end{align}
    In order to fuse $o_{A_i}$ with the corresponding $c_{V_i}$, our initial work~\citep{Sterpu2020a} used a learnable transformation taking as input the concatenation of the two vector representations:
    \begin{align}
        o_{AV_i} & = W_{AV} [o_{A_i}; c_{V_i}] + b_{AV} \label{eq:oavi}\\
    \nonumber &  \mathsf{where} \ W_{AV}  \in  \mathbb{R}^{\mathrm{h\,x\,2h}}, b_{AV} \in \mathbb{R}^{\mathrm{h}}
    \end{align}
    Later, with a Transformer encoder replacing the recurrent one, we chose in \cite{Sterpu2020b} a simpler fusion step without additional learnable parameters:
    \begin{align}
        o_{AV_i} & = o_{A_i} + c_{V_i} \label{eq:fusion2}
    \end{align}
    Since the Audio-Visual Transformer model in \cite{Sterpu2020b} took advantage of the visual modality of speech at a similar rate as the LSTM variant in \cite{Sterpu2020a} compared to an equivalent audio-only model, in this work we will adopt the simpler fusion approach in equation~\eqref{eq:fusion2} with no learnable parameters. This will keep the overall number of parameters to a minimum level, therefore increasing the chance of correct generalisation on a relatively small dataset. 

    AV Align lacks the specification of the temporal limits of integration, which is an inherent limitation of the attention mechanism. Specifically, any fused representation $o_{{AV}_i}$ is conditioned on each visual representation $o_{V_j}$ for any $j$ from 1 to $M$. Consequently, this scheme would no longer allow us to control the encoding latency that was achieved with Taris. In the next section we will address this challenge and will propose an audio-visual extension of Taris based on a windowed version of AV Align.

    \subsection{Audio-Visual Taris}
    \label{sec:avtaris}
    
    The visual modality of speech does not contain sufficient linguistic information to allow the prediction of word boundaries. As a result, we cannot use the same counting strategy as with the audio modality in order to segment visual speech. Learning to segment the audio modality was necessary because the auditory and symbolic modalities of speech exist on different timescales, and we found the concept of words as the linking element between them. However, the audio and video modalities share the same time axis and can be integrated more easily, by only taking into account the different sampling rates. Having prior knowledge of the natural asynchrony between auditory and visual speech allows us to set an upper limit on the audio-visual integration window.
    
    We describe an audiovisual extension of Taris. Given a sequence of visual representations $\mathbf{v} = \{v_1, v_2, \ldots , v_M\}$ corresponding to the audio track $\mathbf{a} = \{a_1, a_2, \ldots , a_{\mathbf{N}}\}$ of the same spoken utterance, we define a symmetrical integration window of length $2B+1$ centred on a visual frame index $j$ and apply a \emph{constrained} cross-modal alignment between modalities:
    \begin{align}
        c_{V_i} & = Attention\ (o_{A_i}, o_{V_{\ j - B\ :\ j + B\ }}) \\
        j & = \Bigl\lfloor (i + 1) \frac{N}{M} \Bigr\rfloor - 1
    \end{align}
    For any audio frame $i$, the index $j$ is calculated as the nearest time-aligned video frame, e.g. audio frame 50 corresponds to the video frame 25 when the audio has twice the sampling rate of the video (i.e. $N = 2M$). Compared to $c_{V_i}$ in Equation (5) from the offline multimodal architecture used in our prior work \citep{Sterpu2020a}, here the alignment is performed within a window of $2B+1$ visual frames, which is only a fraction of the full length $M$ of the visual sequence. Consequently, an audio representation only depends on temporally local video representations, preserving the eager decoding property of Taris. The audio and visual representations are integrated as following:
    \begin{align}
        o_{AV} = c_V + o_A
    \end{align}
    To complete the online audio-visual design, we only need to predict the gating signal $\alpha_i$ from the fused representations $o_{AV}$ instead of the audio ones seen in equation~\eqref{eq:alphai}:
    \begin{align}
        \alpha_i = \mathsf{sigmoid}(o_{AV_i} W_G  + b_G)
        \label{eq:c5:alphaiav}
    \end{align}
    This strategy allows us to investigate if AV Taris can learn to count words in fluent speech better from the fused audiovisual representations instead of the audio ones alone.

    \subsection{Latency analysis}
    
     Calculating the delay between the first audio frame timestamp and the first output unit is non trivial and depends on several factors.
     
     First, the encoding look-ahead and look-back parameters $e_{LA}$ and $e_{LB}$ define the receptive field in audio frames of a learnt audio representation $o_{A_i}$. The absolute encoding delay is a function of $e_{LA}$, the number of layers in the Transformer network, and the audio frame duration. We provide in Section~\ref{recfield} a more precise measurement of the encoding latency.
     
     Second, each audio representation $o_{A_i}$ requires up to B future frames from the visual modality for the multimodal re-contextualisation step. The larger value between the receptive field of the self-attention operation in the last layer of the audio encoder and B will give the encoding latency of the multimodal system.
     
     Third, the decoding look-ahead parameter $d_{LA}$ defines the number of audio segments required to start decoding the first grapheme in an output word unit. Although this number of segments is constant, the frame length of $d_{LA}$ segments is dynamic and context dependent. In other terms, the first grapheme in the next word can be decoded once the cumulative sum of $\alpha_i$ becomes greater or equal to $d_{LA}$.
     
     Finally, we note that an alternative decoding approach in Taris is to gradually increase the segment look-ahead from 0 to $d_{LA}$, and consequently to provide up to $d_{LA} + 1$ updates for the same word. This can be more practical when an immediate, less accurate transcription is needed, accepting that it is subject to corrections depending on the future context.

    \subsection{Complexity analysis}
    When operating on a single modality, Taris requires a negligible overhead in parameters and operations over the original Transformer. The only extra parameters are given by the $W_G$ and $b_G$ variables in equation~\eqref{eq:alphai}, which amount for $h+1$ scalars in the total model size. Equations \eqref{eq:alphai}-\eqref{eq:valid} describe the additional operators mainly consisting of a matrix vector multiplication followed by a sigmoid activation for every audio frame, and the update of a scalar cumulative sum. Since attention masking is already performed by the original Transformer to take into account the true input length in a minibatch, equation~\eqref{eq:tarisctx} does not represent an overhead, and the only additional operation needed at each decoded timestep is the computation of the segment mask $V$ in equation~\eqref{eq:valid}. In training, this mask is computed only once per batch, since we have access to the full output sequence and know the positions of all the $SPACE$ tokens in advance. The mask $V$ is directly applicable on the tensor product performed by the Transformer architecture between the queries and keys by adding a large negative value outside the mask before applying the softmax operation. This grants Taris a highly efficient computation strategy that integrates with the Transformer.
    
    On the other hand, AV Taris additionally requires the application of a Convolutional Neural Network (CNN) on the input images, and an encoder of the same nature as the one used for the audio representations. A thorough discussion on the computational complexity of various CNN architectures is out of the scope of this article. Nevertheless, the overall strategy we propose is invariant of the specific CNN and sequence encoders options, which may be optimised with respect to the requirements of the intended application.
    
    \subsection{Considered alternatives}
    
    Attention-based sequence to sequence models learn an explicit alignment between the output tokens and the speech frames. We initially considered leveraging the alignments corresponding to each SPACE token of a pre-trained offline model, and using this as a supervision signal to train the gate $\alpha_i$ from equation~\eqref{eq:alphai}. However, visually inspecting these alignments revealed that there is no clear delimitation between the spoken words in English, with the softmax weights not being skewed towards a low number of frames. This observation is in line with our intuition that the SPACE token rarely corresponds to a short pause in English speech, and instead has a more analytic role which demands its inference from the acoustic differences between several words, or from the intrinsic language model in the decoder. In effect, such alignment information would only represent a crude approximation of the true boundaries between the spoken words, and it would be a very noisy supervision signal for our gate $\alpha_i$ to learn.
    
    In Taris, we can choose to constrain the output of the gating unit $\alpha_i$ in equation~\eqref{eq:alphai} to follow a specific distribution. For example, \citet{Hou2020} train their gating unit to follow a Bernoulli distribution, making values very close to 0, or very close to 1, more likely. During our initial experiments with a scaled sigmoid function (i.e. \mbox{$1 / (1+exp(-kx)), k>1$}), we noticed that this unit does not typically have values close to the extremities of the range, and achieves a slightly higher word counting loss than standard unscaled sigmoid. We speculate that the gating unit learns to \emph{accumulate} cues at the sub-word level in order to solve the word counting task, and an eventual binary output behaviour may only be feasible when coupled with a recurrent process to keep track of an internal state, which in turn would complicate the design. In our work, $\alpha_i$ is predicted directly from the hidden state $o_{A_i}$ with a feed-forward neural network.
    
    As an alternative to the sigmoid activation we also considered the hyperbolic tangent function, which has an output range between $(-1..1)$. The negative output values have the potential to enable a broader range of word counting strategies, such as assigning higher confidence scores and eventually correcting them later based on future evidence. With the sigmoid activation, the system does not have the opportunity to make corrections and has to adopt a more defensive approach. On the other hand, a sufficiently large receptive field may reduce the need for such corrections. In our initial experiments we did not see a significantly improved word loss with the $tanh$ activation, and, since Taris is a relatively new model, we decided to apply the law of parsimony and maintain the $sigmoid$ until empirical evidence demands otherwise.

    \subsection{Comparison to related work}
    
    \cite{Dong2020} categorise end-to-end speech recognition models into label-synchronous and frame-synchronous models. The former refers to models that derive the contextual acoustic units from the soft alignment with the state of an auto-regressive decoder receiving grapheme labels as inputs. In contrast, the latter derive acoustic labels directly from the audio representations by removing the decoder, and thereby do not model the conditional dependence between the labels.
    
    Taris is more closely related to the label-synchronous class, as it maintains a soft attention mechanism between the decoder and the encoder. However, Taris derives a segmentation signal directly from the audio representations, and the soft alignment is only allowed within a well defined dynamic window. This contrasts with the model proposed by~\citet{Dong2019}, which also predicts a normalised weight per frame, but uses these weights directly once they sum up to approximately 1.0 to linearly combine the corresponding audio representations into a single state from which the segment label is estimated. An approach similar to the one of~\cite{Dong2019} was previously introduced in~\cite{Li2019}, however they only test the method on Mandarin speech. \cite{Li2019} anticipate problems on languages such as English with less clear boundaries between linguistic units and complex orthographies.
    
    Taris is more closely related to the approach of~\citet{Hou2020} performing segment level attention. However, \citet{Hou2020} take a different approach to train the boundary detection unit by sampling from a Bernoulli distribution, which makes the model non-differentiable, and resort to policy gradients. Experimentally, they find that the cumulative sum of the boundary unit requires a dynamic threshold ranging from 0.2 to 0.55 for optimal decoding performance. This suggests that the approach we take with Taris not enforcing a specific distribution on the output of the gating unit, and only requiring the total sum to be close to the word count, is likely enabling the learning of a more flexible counting mechanism. The sigmoidal unit in Taris does not enforce the notion of a hard boundary, but instead we design the decoder to analyse a limited acoustic range covered by the cumulative sum of the gating unit.
    
    \cite{Wang2020} present a related method that uses of a dynamic encoding context to update the audio representations, as does Taris. Our methods differ in the strategy used to detect the boundaries of the dynamic context. \cite{Wang2020} propose a \emph{scout network} to scan for possible word boundaries. Their approach is similar to \cite{Hou2020} or \cite{Li2019}, as it aims to classify each audio frame as either a word boundary or non-boundary. Instead of training the scout network with a single objective function as in the prior work, \cite{Wang2020} see it as a supervised task and make use of forced alignment to provide frame-level word boundary labels. This creates the requirement of a pre-trained forced aligner model, which increases the complexity of the training pipeline. It also represents a possible source of noise when the forced aligner produces noisy or uncertain labels. Instead, our method Taris does not require external labels to train the gating unit, beyond the trivial word count of the considered sentence. Moreover, Taris does not enforce the notion of a word boundary in its encoder, and allows the gating unit to take any intermediate value in the [0..1) range. It is only the decoder of Taris that constructs segments by analysing the cumulative sum of the gate output. We believe these are important advantages of Taris that allow a very simple training. On the other hand, without boundary supervision, Taris does not guarantee that it can learn word boundaries. Instead, the reliability of its boundary predictions is highly coupled with the sentence diversity of the training corpus.
    
    In the space of audio-visual speech recognition, another model that is capable of decoding online is the work of \cite{Makino2019}. Their model is based on the RNN-T, and uses a simpler direct feature fusion strategy for audio-visual integration. They report tailoring the audio feature extraction strategy to match the video frame rate by shifting the spectral analysis window with variable increments. Instead, AV Taris uses the seq2seq architecture as a back-end for sequence modelling, avoiding the training difficulties specific to the RNN-T. Furthermore, we leverage the cross-modal fusion strategy AV Align multimodal feature integration, which decouples the sampling rates of the two input streams.
    
    

\section{Experiments and Results}

We conduct our experiments on the unconstrained audio-visual English speech dataset LRS2~\citep{lrs2}. The main training set of LRS2, used in this work, contains 45,839 spoken sentences, and the test set contains 1243 sentences.
We use the same dataset partitioning, the same audio features, and the same strategy for corruption with additive cafeteria noise at a SNR of 10db, 0db, and -5dB as in \cite{Sterpu2020c} to enable a direct comparison between Taris and AV Taris.
Our implementation of Taris forks the official Transformer model in TensorFlow 2~\citep{tf_transformer}. 

\subsection{Input pre-processing}
\label{exp:preproc}

Our system takes auditory and visual input concurrently. The \textbf{audio} input is the raw waveform signal of an entire sentence. The \textbf{visual} stream consists of video frame sequences, centred on the speaker's face, which correspond to the audio track. We use the OpenFace toolkit of \cite{Baltrusaitis2018} to detect and align the faces, then we crop around the lip region. Complete details of the pre-processing of each stream now follow.

\textbf{Audio input.} The audio waveforms in LRS2 have a sampling rate of 16,000 Hz. The audio signals are additively mixed with cafeteria noise at different Signal to Noise Ratios (SNR) as explained in Section~\ref{sec:training_proc}. We compute the log magnitude spectrogram of the input, choosing a frame length of 25ms with 10ms stride and 1024 frequency bins for the Short-time Fourier Transform (STFT), and a frequency range from 80Hz to 11,025Hz with 30 bins for the mel scale warp. We stack the features of 8 consecutive STFT frames into a larger window, leading to an audio feature vector $a_i$ of size 240, and we shift this window right by 3 frames, thus attaining an overlap of 5 frames between consecutive audio windows.

\textbf{Visual input.} We down-sample the 3-channel RGB images of the lip regions to 36x36 pixels. A ResNet CNN \citep{He2016} processes the images to produce a feature vector $v_j$ of \textbf{128 units} per frame. The details of the architecture are presented in Table~\ref{tab:resnet_details}.

\begin{table}[ht]
\centering
\caption{CNN Architecture. All convolutions use 3x3 kernels, except the final one. The Residual Block \citep{He2016} is in its \emph{full preactivation} variant.}
\label{tab:resnet_details}
\begin{tabular}{rcr}
\textbf{layer} & \textbf{operation}                                      & \textbf{output shape} \\ \hline
0              & Rescale [-1 ... +1]                                                & 36x36x\textbf{3}               \\ 
1              & Conv                                                    & 36x36x\textbf{8}               \\ 
2-3            & Res block                                               & 36x36x\textbf{8}               \\ 
4-5            & Res block                                               & 18x18x\textbf{16}              \\ 
6-7            & Res block                                               & 9x9x\textbf{32}                \\ 
8-9            & Res block                                               & 5x5x\textbf{64}                \\ 
10             & Conv 5x5                                                & 1x1x\textbf{256}
\end{tabular}
\end{table}

\subsection{Training procedure}
\label{sec:training_proc}

The acoustic modality is corrupted with only \emph{Cafeteria} noise, as this noise type was found the most challenging in \cite{Sterpu2018b}, and the noise source did not influence the conclusions. We train our systems in four stages, first on clean speech, then with a Signal to Noise Ratio (SNR) of 10db, 0db and finally -5db.
Each time we increment the noise level we also copy the model parameters rather than train from scratch, speeding up the system's convergence. We train our models on LRS2 for a total of 120 epochs at an initial learning rate of 0.001, decayed to 0.0001 after 100 epochs, on each noise level. The systems are evaluated at the end of the 120 iterations. The training time of AV Taris is approximately 450 seconds for a single epoch of LRS2 on an Nvidia Titan XP GPU.

For the reasons explained in \cite{Sterpu2020c}, we do not use the common End-of-Sentence (EOS) token to pad our target sequences when training our online models.

\subsection{Neural network details}
Our models use 6 layers in the Encoder and Decoder stacks of the Transformer, with a hidden model size $d_{model} \equiv h = 256$, a filter size $d_{ff} = 256$, one attention head, and 0.1 dropout on all attention weights and feedforward activations. The audio-visual models occupy 36 MB on disk, and are considerably smaller than the typical size of state-of-the-art models used in benchmarks. We chose this model size so we could train it on a single GPU of 12 GB of memory with a minibatch size of 32. We presume that a larger model may bring a similar level of improvement to both the online and the offline systems if we wanted to pursue a better absolute accuracy. This would come at the cost of slower, more expensive training iterations.

\subsection{Analysis of the receptive field}
\label{recfield}

As we described our data pre-processing setup in Section~\ref{exp:preproc}, one audio frame is obtained by stacking 8 STFT frames taken over 25ms windows with 10ms strides. Each new audio frame includes the previous 5 STFT frames, so the additional non-overlapping information is represented by 3 STFT frames. In greater detail, the first audio frame achieves an effective range from 0ms to 95ms. The second audio frame starts at 30ms going up to 125ms, followed by the third frame from 60ms to 155ms, and so on.

The first layer in our Transformer encoder network has a receptive field in frames controlled by the $e_{LA}$ and $e_{LA}$ parameters. We preserve the same mask throughout the entire Transformer stack. This means that the superior layers can access a broader receptive field with respect with the audio input. A representation at position $k$ in the Transformer layer $l$ is then indirectly conditioned on the audio input up to the position $k + l\cdot e_{LA}$. We leave the fine tuning of this connectivity design for latency optimisation as future work.

\subsection{Learning to count words in audio-visual speech}
\label{exp:avcounting}

\begin{figure}[t]
        \centering
        \subfloat[Character Error Rate]{
        \includegraphics[width=0.5\linewidth]{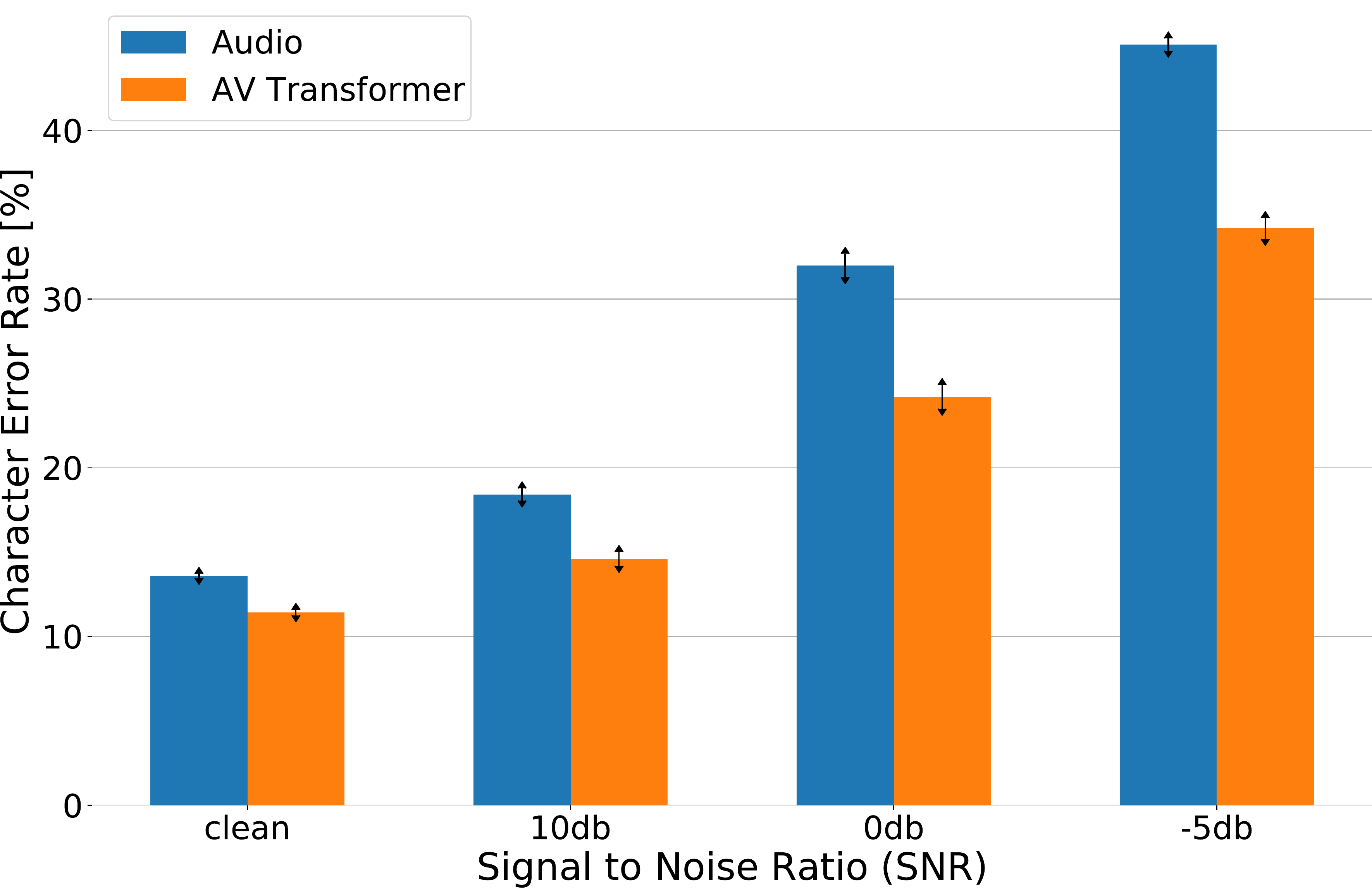}
           \label{fig:avcounting_cer}
        }
        \subfloat[Word Count Loss]{
        \includegraphics[width=0.5\linewidth]{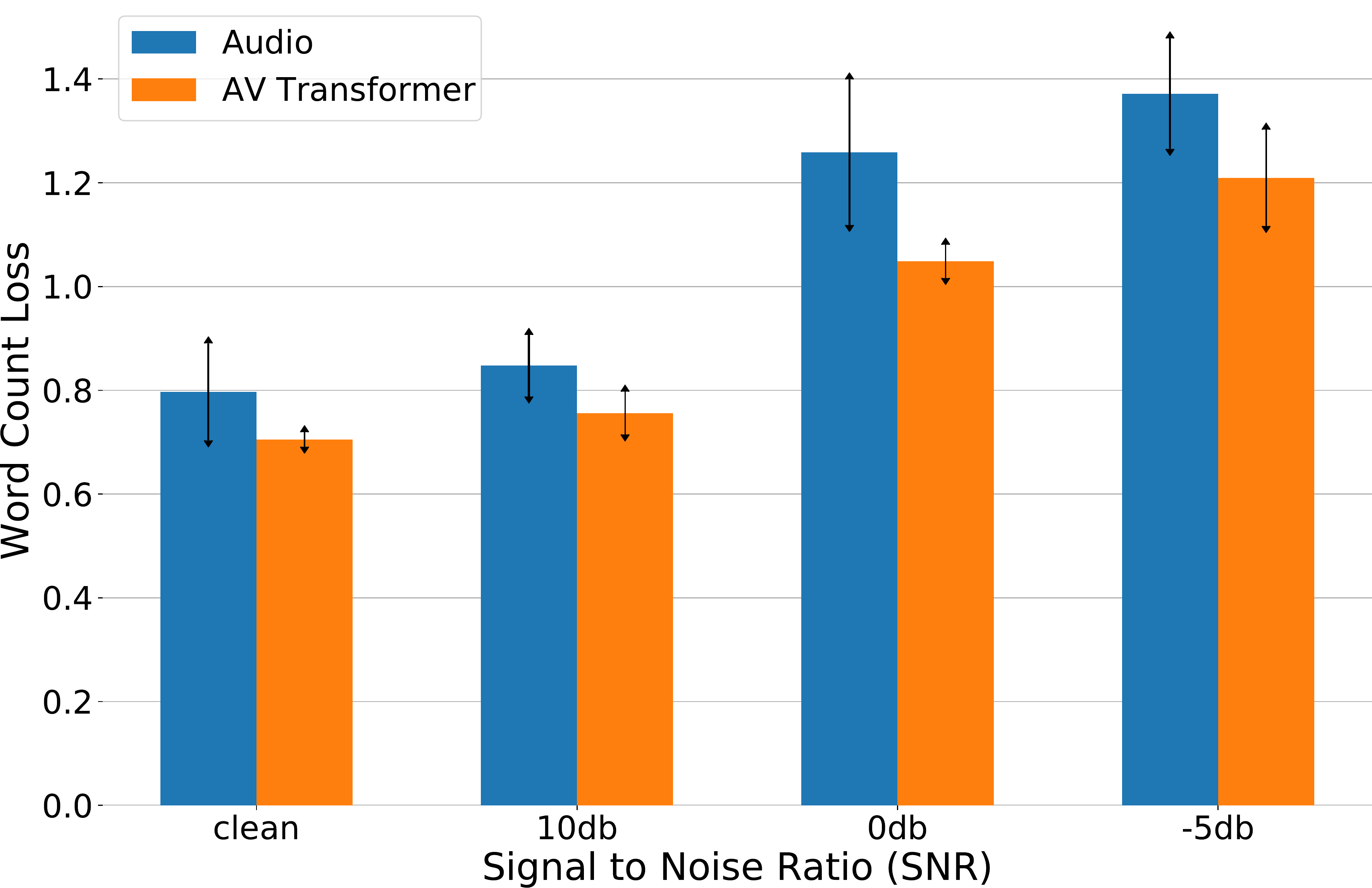}
           \label{fig:avcounting_wcl}
        }
        \caption{Evaluation of the offline Audio and the Audio-Visual Transformer on LRS2 with the word counting loss enabled}
        \label{fig:avcounting}
    \end{figure}

We are interested in studying if the word count in fluent speech can be estimated with a higher accuracy from audio-visual cues than from the audio modality alone. In \cite{Sterpu2020c} we saw that the encoding look-ahead length does not have a major influence on either the word counting error or the character error rate. Therefore, in this experiment we limit our analysis to counting words from audio-visual representations with the offline models having infinite context available. We train Audio and Audio-Visual Transformer models on LRS2 and repeat the experiment for five different random initialisations. We plot the average Character Error Rate and the Word count loss of the two systems in Figure~\ref{fig:avcounting}. The arrows indicate the standard deviation across the five trials. To help the visual network learn good representations, the Audio-Visual Transformer uses the auxiliary Action Unit loss described in \cite{Sterpu2020a}. Note that the Transformer-based models used in this experiment take advantage of the entire input sequence when updating each representation, and cannot be used in an online setting.

From Figure~\ref{fig:avcounting_wcl} it can be seen that the average word count loss of the Audio-Visual Transformer is slightly lower than the one of the Audio model, while the recognition accuracy shown in Figure~\ref{fig:avcounting_cer} stays approximately the same as in our prior work \cite{Sterpu2020b}, where we did not use the Word Loss. This aspect suggests that the visual cues may be informative of word boundaries in fluent speech, although it is difficult to draw conclusions regarding the statistical significance of this result from only 5 trials.

\subsection{Online audio-visual decoding}

The decoder in our previous experiment had access to the entire encoder memory. For our online models in this section we opt for an encoder look-ahead $e_{LA}$ of 11 frames and infinite look-back $e_{LB} = \infty$, since we showed in \cite{Sterpu2020c} that there are diminishing gains beyond this threshold. In that work we have also demonstrated that Taris can leverage the gating signal $\alpha$ to limit the dynamic range of decoder-encoder attention and still match the error rate of the offline Transformer. We now investigate how the audio-visual extension of Taris compares to the offline Audio-Visual Transformer.

In Section~\ref{exp:avcounting} we have seen that an Audio-Visual Transformer with infinite look-back and look-ahead encoding context achieves a slightly lower word counting cost than an Audio-only counterpart. Therefore, learning to count from the fused audio-visual representation does not degrade the word counting accuracy. The system may additionally take advantage of the visual modality to further improve its counting estimate.
 
\begin{figure}[t]
        \centering
        \subfloat[Character Error Rate]{
        \includegraphics[width=0.5\linewidth]{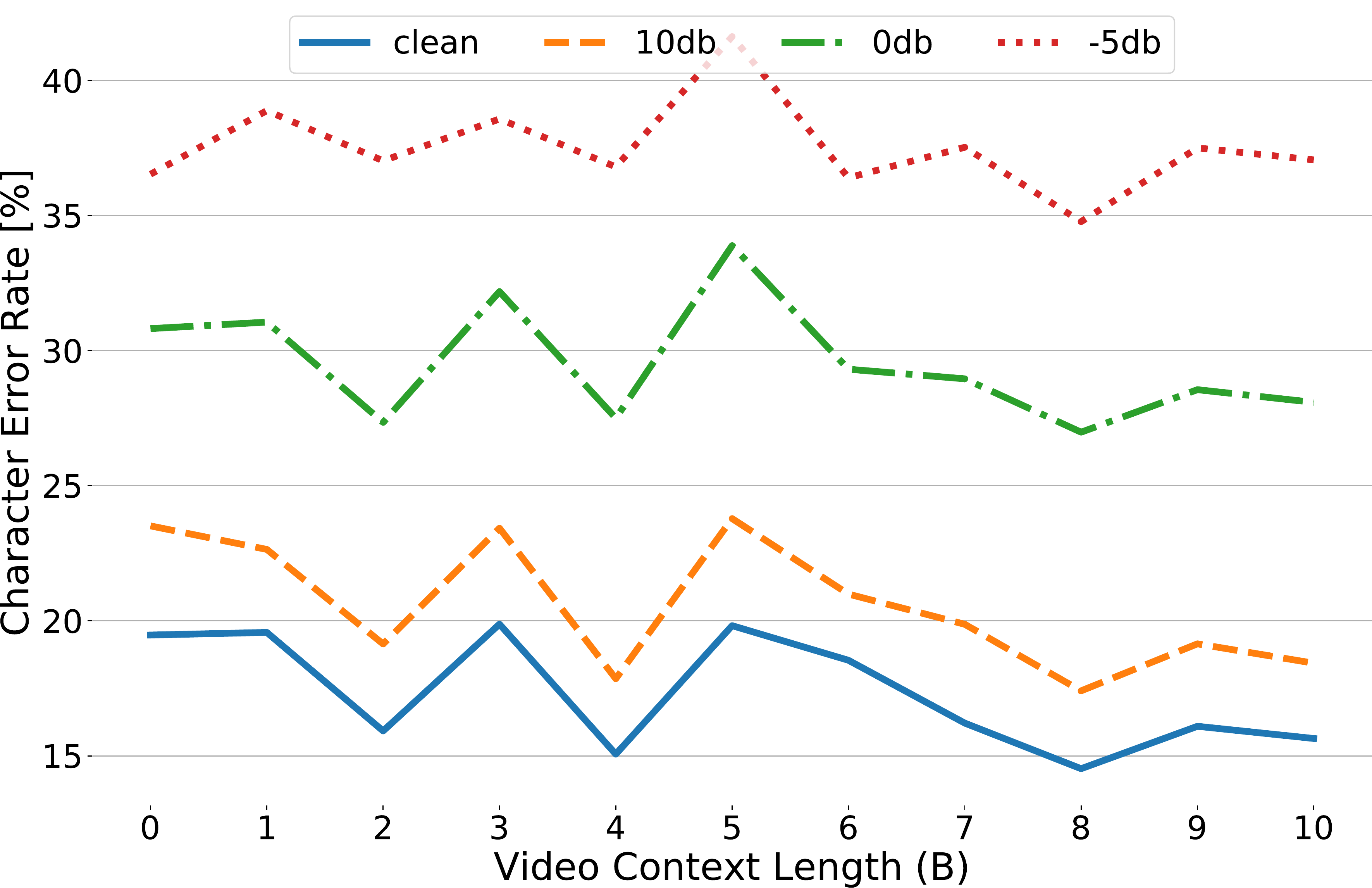}
           \label{fig:avtaris_cer}
        }
        \subfloat[Word Counting Loss]{
        \includegraphics[width=0.5\linewidth]{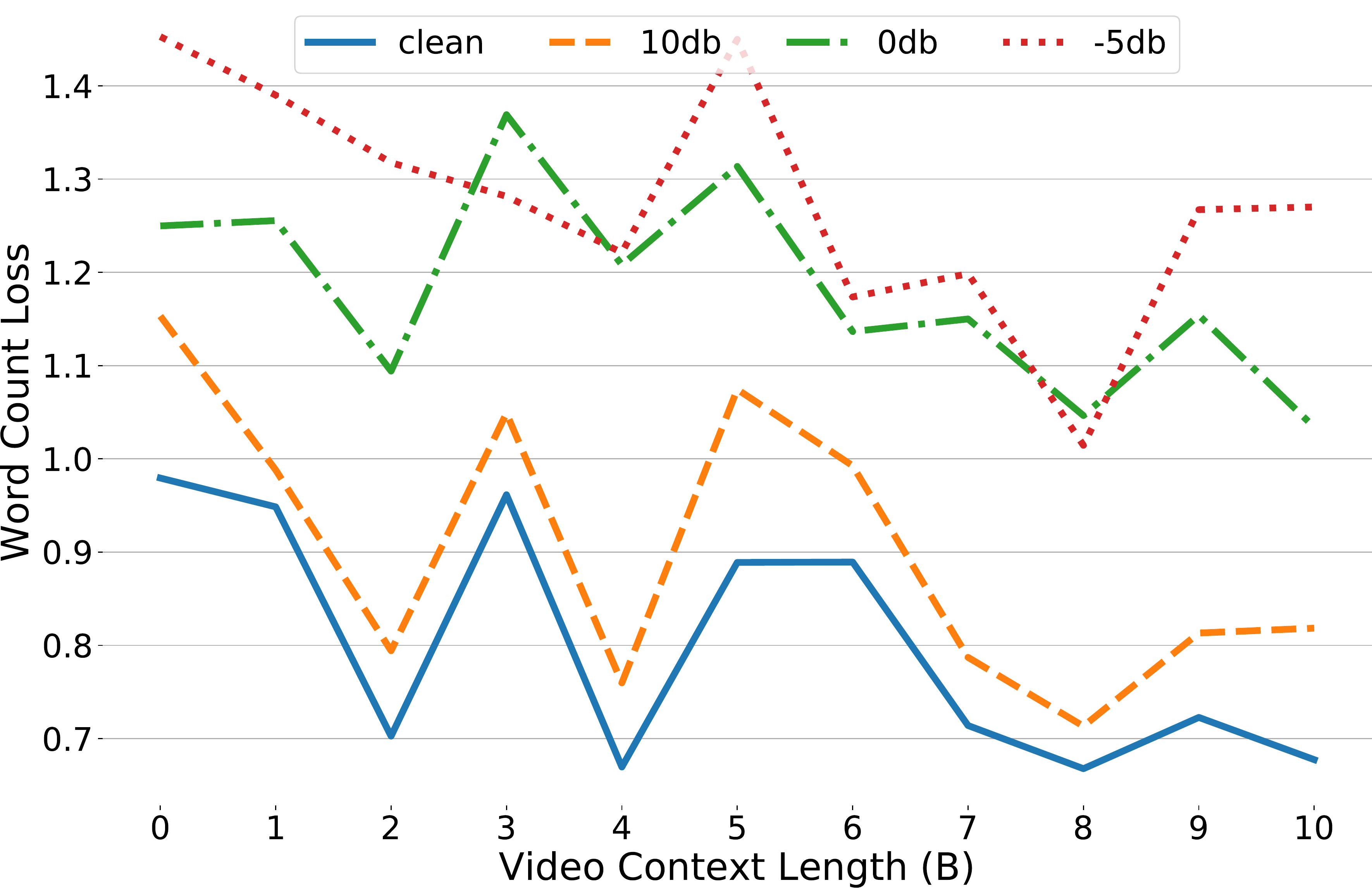}
           \label{fig:avtaris_wcl}
        }
        \caption{Evaluation of AV Taris on LRS2 when varying the size of the symmetrical window used for the soft-selection of the visual representation aligned with each audio representation}
        \label{fig:avtaris}
    \end{figure}

 We evaluate the AV Taris model for an increasing length of the cross-modal attention window, controlled by the length parameter $B$. The window length is defined as $len(w) = 2 B + 1$, as it extends symmetrically in both directions. When $B=0$ the system is similar to a down-sampled version of feature fusion that bypasses the requirement to have identical sampling rates for both modalities. For $B>0$, the window is extended with $B$ frames to the left and to the right respectively. The results obtained with AV Taris displayed in Figure~\ref{fig:avtaris}.
 
Our best AV Taris systems are obtained when B is equal to 2, 4, and greater than 7. Given the difficulty of evaluating a neural network with respect to all possible source of variations such as weight initialisation, example shuffling or random dropout, which remains an open research question (see \cite{Dror2019} for a thorough analysis), it is impractical to perform a proper significance test for the systems displayed in Figure~\ref{fig:avtaris}. One notable trend is that the error rates stabilise with the increasing length of the video attention window. Beyond 7 future video frames there are diminishing returns in decoding accuracy, although it may be possible to further restrict the context to only 2 video frames and still achieve comparable results.

We observe an absolute difference of approximately 3\% at all the noise levels between AV Taris and the offline AV Align Transformer  model shown in Figure~\ref{fig:avcounting_cer}.
AV Taris is still superior to the audio-only models both in their offline and online variants at higher levels of noise. However, in low noise conditions up to 10db, the degradation of AV Taris over the AV Align Transformer considerably diminishes the benefit of cross-modal alignment over audio-only modelling. Nevertheless, the contribution of the visual modality in clean speech may span beyond the average Character Error Rate metric used in this work. This remains a topic for further investigation.

\subsection{Segment length histograms}

As in \cite{Sterpu2020c}, we want to analyse the typical lengths of the segments that AV Taris discovers. Although it would be much stronger to investigate the correlation between segments and words more explicitly, beyond duration,
we appreciate there are several drawbacks for doing so. First, Taris does not explicitly optimise the correct segmentation of words from a spoken utterance, but this is only a by-product of the Character Error Rate objective. Furthermore, since Taris allows left and right context when decoding a word, this diminishes the necessity for precise word boundaries. Instead, ensuring a reasonable duration for the segments Taris discovers is essential to low latency decoding, which is the main objective of our work.

\begin{figure}[t]
    \centering
    \subfloat[Audio Taris clean]{\includegraphics[width=0.33\linewidth]{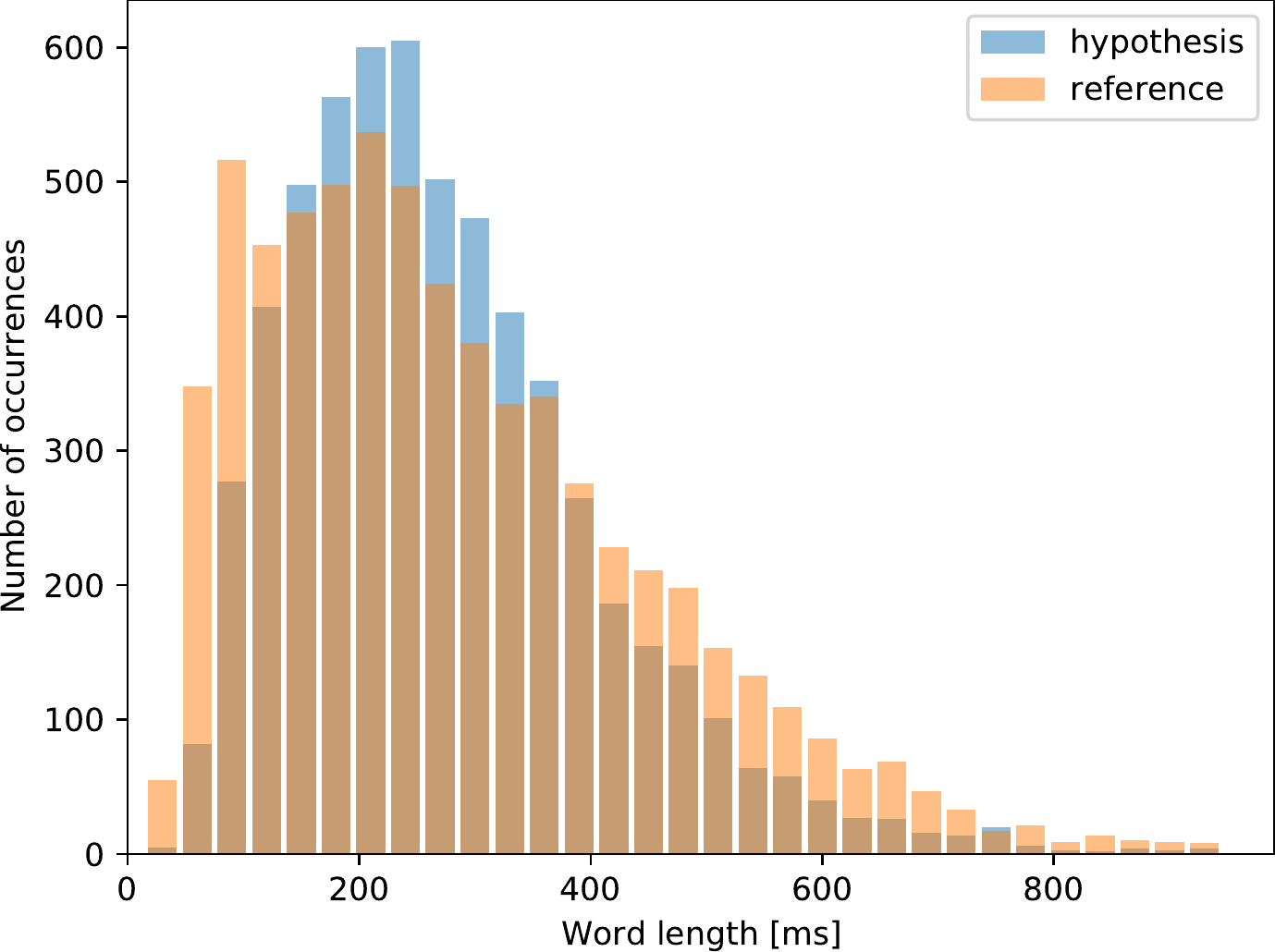}
    \label{fig:hist:taris}}
    \subfloat[AV Taris clean]{
    \includegraphics[width=0.33\linewidth]{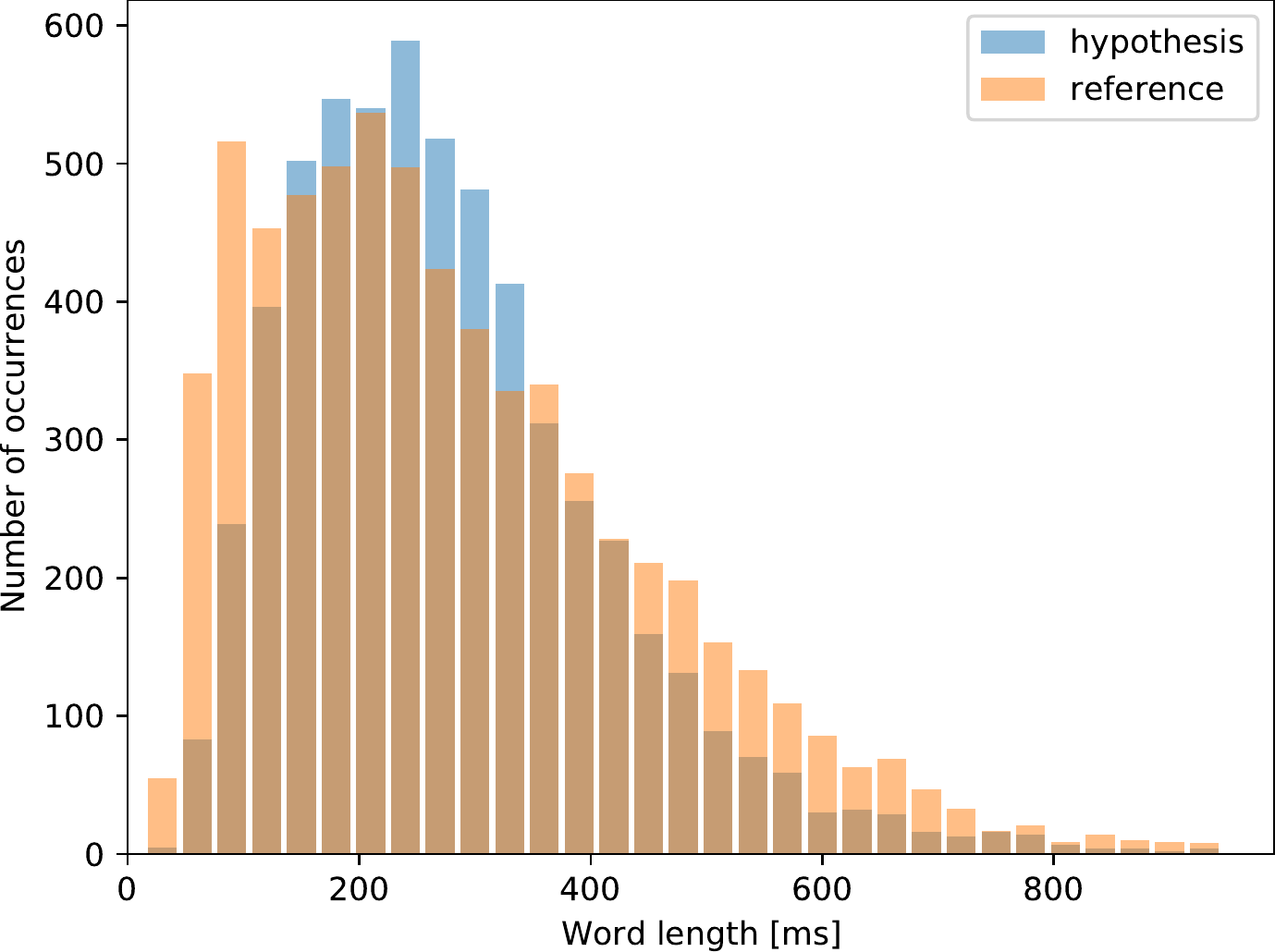}
    \label{fig:hist:avtaris}}
    \subfloat[AV Taris -5db]{\includegraphics[width=0.33\linewidth]{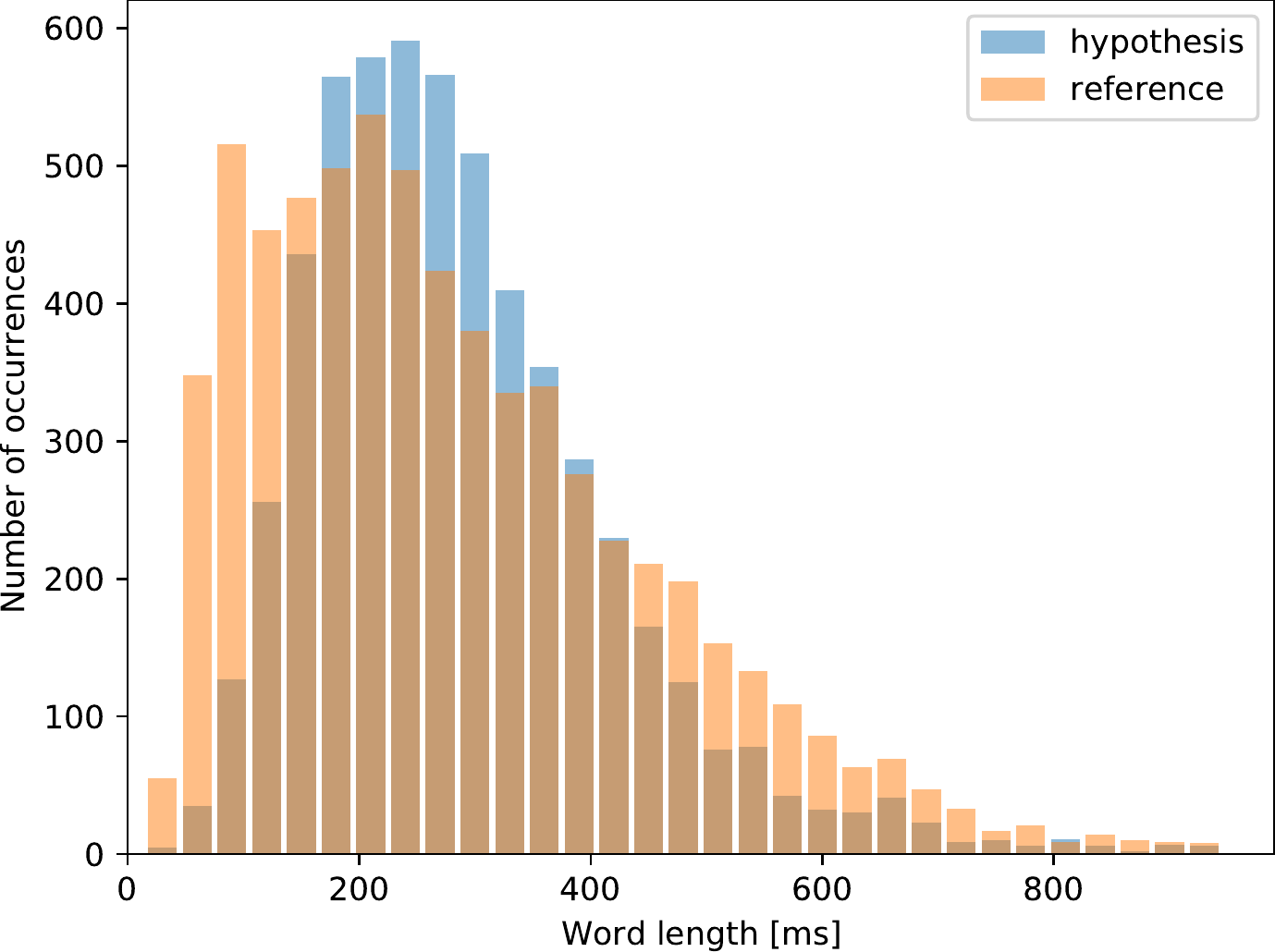}
    \label{fig:hist:avtaris-noisy}}
    \caption{The histograms of the }
    \label{fig:histograms}
\end{figure}

In Figure~\ref{fig:histograms} we plot the histograms of the segment lengths, revealing the audio context used during the decoding process. To compute the length of a segment, we first detect those timestamps where the cumulative sum of the gating signal $\alpha_i$ passed to an integer value. Next, we compute the difference between any two consecutive crossing points. We then transform these values representing frame counts to milliseconds by multiplying them with the same amount that was used to shift the audio feature window to the right when pre-processing the audio data, namely 30ms. Finally, we compute the histograms of all the segment lengths from all the sentences in the test set of LRS2. Figure~\ref{fig:hist:taris} displays the histogram corresponding to Audio Taris on clean speech, whereas Figures~\ref{fig:hist:avtaris}-\ref{fig:hist:avtaris-noisy} show the histograms of AV Taris on clean and noisy speech respectively. In orange, we overlay the histogram of the ground truth / reference word lengths estimated with the Montreal forced aligner tool of \cite{McAuliffe2017}.

In all cases, we notice that the word counting strategy adopted in AV Taris leads to a reasonable approximation of the word length distribution estimated with the forced aligner. At this level of coarseness, it is hard to appreciate the differences in segmentation between Audio Taris and AV Taris on clean speech. The only noticeable difference is between clean and -5db speech, where the mean of the distribution shifts to the right in the latter case. This suggests that the AV Taris system has the tendency to create longer segments on noisy speech by fusing several shorter segments detected on clean speech. The overall effect would be an increased decoding latency on noisy speech, suggesting that the system requires more audio context before becoming sufficiently confident to initiate decoding.
\section{Discussion and conclusion}

We have proposed a multimodal extension of the Taris system as a fully differentiable solution to online audio-visual speech recognition. To accomplish this, we revised the cross-modal attention mechanism in AV Align by limiting the attention span to a fixed window of video representations centred on each audio frame. As a result, we have achieved an audio-visual speech recognition system that can decode online. Experimentally, we found that the accuracy of the audio-visual extension of Taris lags behind the offline Transformer-based AV Align system approximately by an absolute 3\%. The offline model could exploit the entire utterance for both cross-modal alignment and decoding, explaining one possible source for the difference. We believe that the modelling assumptions in both AV Align and Taris are sufficiently general to transfer to other multimodal speech processing tasks.

\cite{Chiu2019} and \cite{ Narayanan2019} report that the neural networks used in speech recognition struggle to generalise to sentences that are considerably longer than the ones seen in training. Instead, Taris only models the the local relationships in speech, and is structurally unaffected by the sentence length at inference beyond the analysis window defined by the look-back and look-ahead parameters. The same property was seen in the Neural Transducer (NT) model of \cite{Jaitly2016}, although Taris allows adaptive segments and simplified training by avoiding the problems introduced by the NT's end-of-block token. Similar to the NT model, Taris repeatedly applies a sequence to sequence model over consecutive audio windows. The NT model processes fixed length blocks, and does not need to \emph{learn} a segmentation. Their mechanism increases the complexity of the optimisation algorithm. More precisely, the model introduces an additional end-of-block token in the output domain that needs to be emitted once per every audio window. This generates the problem of having to search for an optimal alignment in training between the longer sequence of predicted labels containing the additional token and the shorter ground truth sequence. Taris avoids this problem by not making use of end-of-block tokens. Instead, Taris analyses dynamic windows of speech centred on a word of interest. On the other hand, Taris does not guarantee the reliable segmentation of the spoken utterance into words. It only facilitates the compensation of eventual segmentation errors with a controllable number of look-back and look-ahead segments that the decoder is allowed to attend to. Studying the internal segmentation achieved by Taris remains a topic for future exploration.

Compared to alternative online models such as the RNN Transducer, Taris reduces the computational cost of training and the engineering cost of maintaining the hardware-specific software implementation of the RNN-T objective function. Additionally, it springs from the sequence to sequence model architecture that is currently outperforming alternative approaches. We believe that both the audio and the audio-visual variants of Taris represent a step forward for increasing the accessibility of audio-visual speech recognition technology, although they still require validation at a much larger scale than this article could afford. Considering the current limits of ASR technology to accurately decode highly unstructured and noisy speech as seen in the recent CHiME-5 challenge~\citep{Barker2018}, we believe that the original contributions of this work will enliven the adoption of AVSR solutions.

An interesting behaviour of Taris concerns the handling of word contractions, such as \emph{you're, that's, don't, it's, let's}, and others. In our work, we considered that written words are exclusively separated by spaces, as seen in equation~\eqref{c4:eq:space}. Unless there is a systematic error in the transcriptions, Taris has the potential to learn the acoustic differences between "you're" and "you are". The system maintains its own segment counter (the cumulative sum of $\alpha_i$), and has sufficient freedom to decide which form to transcribe. When "you are" is more likely, then the fraction of $\alpha_i$ added to $\Sigma \hat{w}$ may simply be one unit greater than when "you're" is preferred. Taris can also recover from potential errors since it uses a context window larger than a single segment. Depending on the intermediate scores $\alpha_i$, the decoding of "are" in "you are" may then be conditioned on the acoustic representations corresponding to "you" and other adjacent segments.
On the other hand, the general formulation of the word counting task in Taris may be problematic in the case of modelling silences. Since silences are generally not annotated in the human transcriptions, Taris implicitly includes all the audio frames not substantially modifying $\alpha_i$ to the adjacent segment. Consequently, the decoder performs a soft alignment even over those uninformative silence frames. Increasing the efficiency of this process represents a possible direction of improvement.


It is unlikely that humans learn to segment speech by counting words in full sentences. We are not offered the word count in a numeric format as a supervision signal. Why would it be appropriate to design a speech recognition system based on this aspect? We believe there are several reasons. First, this task would not be impossible for humans if it was formulated as a puzzle for finding patterns in a foreign language. Language acquisition in humans involves a long term process of teaching simpler, isolated words before gradually increasing the difficulty. These learning strategies have not fully matured in our machine learning technology. On the other hand, it is very common, and cheap, to produce a speech dataset annotated at the sentence level, without intermediate phone-level or word-level alignments. Therefore we are already asking computers to solve the speech recognition challenge differently from the way we learn a spoken language. We argue that learning to count words is a good compromise with respect to our existing technology and datasets when aiming to segment a spoken utterance.

\section{Acknowledgement}

We would like to thank Christian Saam for the feedback offered on the manuscript.

This research was conducted with the financial support of Science Foundation Ireland under Grant Agreement No. 13/RC/2106 at the ADAPT SFI Research Centre at Trinity College Dublin, Ireland. The ADAPT SFI Centre for Digital Media Technology is funded by Science Foundation Ireland through the SFI Research Centres Programme and is co-funded under the European Regional Development Fund (ERDF) through Grant \# 13/RC/2106. Our work is supported by a Titan Xp GPU grant from NVIDIA.

\bibliography{mybib}

\end{document}